\documentclass[runningheads]{llncs}
\usepackage[T1]{fontenc}
\usepackage{graphicx}
\usepackage{amsmath}
\usepackage{caption}
\usepackage{subcaption}
\usepackage{url}
\usepackage{enumitem}
\usepackage{booktabs}
\usepackage[table]{xcolor}
\usepackage{amssymb}

\begin{document}
\title{SpaER: Learning Spatio-temporal Equivariant Representations for Fetal Brain Motion Tracking}
\titlerunning{SpaER: Efficient Fetal Motion Tracking}
\author{Jian Wang \inst{1}\and
Razieh Faghihpirayesh \inst{1,2} \and
Polina Golland \inst{3} \and
Ali Gholipour\inst{1, 4}}
\authorrunning{Wang et al.}
\institute{ Computational Radiology Laboratory, Boston Children's Hospital, and Harvard Medical School, Boston, MA, USA\and
Department of Electrical Engineering, Northeastern University, Boston, MA, USA \and 
Computer Science and Artificial Intelligence Laboratory (CSAIL), Massachusetts Institute of Technology, Cambridge, MA, USA \and 
Department of Radiological Sciences, University of California Irvine, CA, USA}

\maketitle              
\begin{abstract}
In this paper, we introduce SpaER, a pioneering method for fetal motion tracking that leverages equivariant filters and self-attention mechanisms to effectively learn spatio-temporal representations. Different from conventional approaches that statically estimate fetal brain motions from pairs of images, our method dynamically tracks the rigid movement patterns of the fetal head across temporal and spatial dimensions. 
Specifically, we first develop an equivariant neural network that efficiently learns rigid motion sequences through low-dimensional spatial representations of images. Subsequently, we learn spatio-temporal representations by incorporating time encoding and self-attention neural network layers. This approach allows for the capture of long-term dependencies of fetal brain motion and addresses alignment errors due to contrast changes and severe motion artifacts. Our model also provides a geometric deformation estimation that properly addresses image distortions among all time frames. To the best of our knowledge, our approach is the first to learn spatial-temporal representations via deep neural networks for fetal motion tracking without data augmentation. We validated our model using real fetal echo-planar images with simulated and real motions. Our method carries significant potential value in accurately measuring, tracking, and correcting fetal motion in fetal MRI. 
\end{abstract}

\section{Introduction}

Motion estimation is a crucial process designed to correct image artifacts caused by object motion, particularly in medical fetal imaging. Its applications extend to various processing tasks, including image segmentation~\cite{faghihpirayesh2022deep,gholipour2017normative,keraudren2014automated}, reconstruction~\cite{gholipour2010robust,kuklisova2012reconstruction,xu2023nesvor}, and pose estimation~\cite{xu20203d,salehi2018real}. Effective motion compensation contributes significantly to the accuracy and efficiency of fetal magnetic resonance imaging (MRI)~\cite{malamateniou2013motion}. Fetal motion tracking faces challenges due to the complexity of fetal movements, discrepancies among data frames, and local geometric distortions. 

To address these challenges, progresses have been made when handling fetal motion in slice-to-volume reconstruction (SVR) to reduce errors caused by slice misalignment. Such techniques include a reconstruction model that aligns image stacks based on 2D slice intersections~\cite{kim2009intersection}, a technique for slice-level motion tracking by utilizing the 3D anatomy coverage from simultaneous slices~\cite{marami2019motion}, and total variation regularization using the primal-dual hybrid gradient method~\cite{tourbier2015efficient}. These methods, however, have limitations in capturing large and rapid fetal movements. Later, spatio-temporal representations were utilized for in-utero image tasks~\cite{liao2016temporal,taymourtash2022spatio,turk2017spatiotemporal}. For example, an iterative framework was proposed to optimize over the temporal domain by a designed reconstruction loss term that included low-rank and total variation regularization. While these approaches incorporate spatio-temporal information, they need to undergo the entire iterative optimization process. This renders them impractical for \textit{real-time} motion monitoring tasks that demand high model efficiency and stability, which are critical for on-scanner quality assurance and prospective acquisition planning~\cite{white2010promo}.

Deep learning approaches have significantly improved the overall model performance of motion estimation~\cite{mahendran20173d}. Methods predicting 3D rigid transformations and real-time fetal motion tracking using various designs of deep neural networks have been introduced~\cite{evan2022keymorph,salehi2018real,singh2020deep}. Two models based on convolutional neural networks (CNNs) were presented to learn motion parameters by minimizing the error between the model output and ground truth in a supervised manner. A transformer-based approach was introduced~\cite{vaswani2017attention} to streamline automatic relevance detection between slices~\cite{xu2022svort}. Building on this model, a reconstruction technique was further introduced, incorporating implicit neural representations, resulting in enhanced image reconstruction performance~\cite{xu2023nesvor}. Instead of estimating rigid motions from the original image space, an unsupervised deep learning framework, KeyMorph, was introduced for robust and interpretable multi-modal medical image registration~\cite{evan2022keymorph} via anatomically-consistent key points. Also, an efficient motion tracking model learns low-dimensional spatial means of equivariant representations for pairs of images~\cite{moyer2021equivariant}. The effectiveness of this approach in capturing significant rigid motions is derived from the intrinsic rotation-equivariant nature of equivariant filters~\cite{cohen2016group}.

While the mentioned methods have shown notable performance, they still encounter difficulties in capturing true spatial-temporal relationships due to their fixed image context range and susceptibility to unpredictable motions and distortions, particularly in extensive MRI sequences. This has motivated our exploration of a spatial-temporal model capable of considering the entire sequence, thereby mitigating motion correction errors arising from motion artifacts, contrast changes, and image distortions across different time points. The core contributions of our proposed method are summarized as follows: 
\setlist[itemize,1]{label=$\bullet$, font=\bfseries}
\begin{itemize}
\item Our model is the first predictive approach in dynamically learning the fetal brain movements spatio-temporally through deep equivariant representations without data augmentation.
\item The framework we propose not only outperforms state-of-the-art methods in motion tracking accuracy but also ensures stable convergence of training and fast inference. This feature holds significant clinical value, particularly in real-time, automated fetal head motion tracking and prospective steering systems, where frequent and substantial motions are prevalent.
\item The theoretical framework we have developed lays the groundwork for rigid motion estimation without the need for network retraining for unseen image modalities, as it inherently learns the nature of rigid transformation.
\end{itemize}
\section{Methodology}
In this section, we first begin with a review of recent work by Moyer et al.~\cite{moyer2021equivariant},  which utilizes equivariant filters to accurately estimate rigid transformations between pairs of images. Following this, we introduce our approach for spatio-temporal sequence estimation of rigid transformations. Utilizing deep neural networks, we focus on learning spatio-temporal representations and provide a detailed overview of the proposed model architecture.

\subsection{Rigid Motion Estimation via Equivariant Filters}
Given a source image $I_0$ and a target image $I_1$, rigid motion estimation aims to identify the best translation $\mathcal{T}$ and rotation $\mathcal{R}$ parameters that define a rigid transformation $Q$ by minimizing the distance between the transformed source image and the target image, represented as:
\begin{equation}
\label{eq: dist} 
E(Q) =  \text{Dist} [I_0 \circ Q , I_1] ,
\end{equation} 
where $\circ$ denotes the composition operator that resamples $I_0$ using the rigid transformation. When this operator is applied to any vector $\textbf{v}$, it yields a transformed vector $Q(\textbf{v})$ of the form $Q(\textbf{v}) = \mathcal{R} \textbf{v} + \mathcal{T}$. Here, $\mathcal{R}^T = \mathcal{R}^{-1}$ indicating that $\mathcal{R}$ is an orthogonal matrix.  

Instead of directly estimating the transformation function $Q$ within the original image space~\cite{salehi2018real,singh2020deep}, recent advancements have introduced an efficient method to compute $Q$ by calculating the spatial means $x^k$ of images~\cite{moyer2021equivariant}. This process involves the application of reduced non-negative equivariant filters $f$ on the images, formulated as:
\begin{equation}
\label{eq: eqfilter} 
x^k = \frac{1}{M_1 M_2 M_3} \sum \gamma f_k (\gamma),
\end{equation} 
where $M_1 \times M_2 \times M_3$ denote the image dimensions, $\gamma$ refers to the image coordinates, and $f_k(\gamma)$ is the value of the $k^{th}$ equivariant filter at $\gamma$. Here, a convolutional filter bank $\mathcal{F} : \mathcal{I} \rightarrow \mathcal{I}(\mathbb{R}^+)^K$ with $K$ non-negative real-valued channels is considered
equivariant under rigid transformations if each channel $\mathcal{F}_k$ satisfies the equivariance property. Such filter banks are constructed by stacking alternating layers of equivariant convolutions and their corresponding non-linearities. Please refer to the formulations of rotation equivariant filters~\cite{moyer2021equivariant,weiler20183d} for more details.  For an image pair $I_0$, and $I_1$, the closed-form update of the rigid transformation can be computed by such low-dimensional spatial means $x_k^{I_1}  x_k^{I_{0}}$ directly. 

\subsection{Spatio-temporal Rigid Motion Estimation}
Given an image sequence $\pmb{I} = \{ I_0, I_1, ... , I_T \}$ changing over time $\pmb{t} =\{ 0, 1, ... , T \}$, we develop an optimization objective for spatio-temporal motion tracking to find the optimal path $\textbf{Q}$ of rigid movement as follows, 
\begin{equation}
\label{eq: spaerenergy} 
E(\textbf{Q}) =  \sum^{T}_{t=0} \text{Dist} [I_t \circ Q_t(z_t, z_{t+1})  , I_{t+1}] ,
\end{equation} 
where $z_t$ represents the spatio-temporal representations across all time frames. Here $\text{Dist} [\ \cdot \ ]$ denotes a distance term that measures the dissimilarity (e.g., sum-of-squared differences, normalized cross-correlation~\cite{avants2008symmetric}, or mutual information~\cite{wells1996multi}) between the aligned and the target sequence. 

We derive the closed-form solution for both translation and rotation parameters between time point $t$ and $t+1$ to characterize $Q_t$,
\begin{equation}
\label{eq: spaercloseform} 
\mathcal{T}_t = z_{t+1} - \mathcal{R}_t z_{t},\quad \mathcal {R}_t = V_t \cdot {U_t}^T,\quad \text{s.t.} \, \, \det (\mathcal{R}_t) = 1,
\end{equation}
where $ {U}_t \Sigma_t V^*_t = z_{t} \cdot z_{t+1}^{T}$, $U_t$ and $V^*_t$ are real orthogonal matrices, $\Sigma_t$ is a diagonal matrix with non-negative real numbers on the diagonal. Setting the determinant of $\mathcal{R}_t$ to $1$ guarantees that it accurately reflects a rigid transformation.

\subsection{Network Design and Training}
\begin{figure}[!bt]
\centering
 \includegraphics[width=1.00\textwidth]{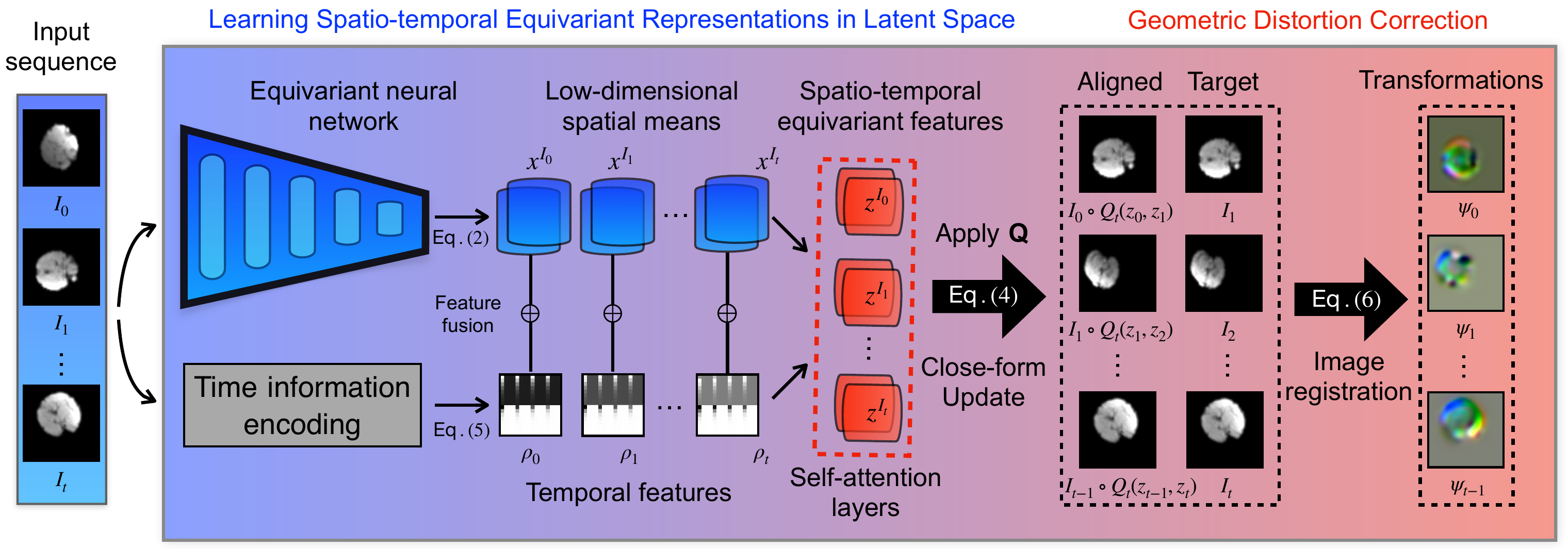}
     \caption{An illustration of the network architecture of our proposed spatio-temporal learning framework, SpaER. Left to right: input data, networks that encode both the temporal and spatial features to learn rigid motions, and the output aligned sequence with a joint geometric distortion correction module. } 
\label{fig: spaer}            
\end{figure}

We develop a deep learning model that is specifically architected to estimate rigid motions as expressed in  Eq.~\eqref{eq: spaerenergy}. Our framework consists of three sub-modules: i) a boosted rigid motion correction neural network that is parameterized by equivariant filters, ii) a temporal encoding module that incorporates time information of the sequence to the equivariant features, and iii) a self-attention neural network that learns the feature correspondence across different time points after taking both features from i and ii. Our framework is illustrated in Fig.~\ref{fig: spaer}. In the following sections, we provide a detailed description of our network architecture and the formulation of our network loss.

\paragraph*{\bf Motion tracking via Equivariant Neural Network}
Let $\Theta$ represent the parameters of the equivariant filters that output spatial means from image spaces. Our model incorporates a 7-layer equivariant neural network designed to capture rotation-equivariant representations through the use of  3D steerable CNNs~\cite{weiler20183d}. 
\paragraph*{\bf Spatio-temporal encoding } 
We employ a temporal encoding scheme that is similar to ~\cite{vaswani2017attention} that maps the time information of input sequence to image features. Suppose that the input representations of spatial means $x \in \mathbb{R}^{T \times d} $ contains the $d$-dimensional spatial means of a sequence (size as $T$). The spatio-temporal encoding outputs $\pmb {\rho} $ using a positional embedding matrix of the same shape $\pmb{\rho} \in \mathbb{R}^{T \times d} $ , whose element on the $t^{\text{th}}$ position is, 
\begin{align}
\label{eq: tempencod} 
\pmb{\rho} (t, 2i) = \sin (\alpha \cdot t ), \quad \quad \pmb{\rho} (t, 2i+1) = \cos (\alpha \cdot t),
\end{align}
where $\alpha = 10^{-8t/d}$ and $i$ is an indicator that is used for making alternate even and odd sequences and $d$ is the dimensionality of the spatial means. 
\paragraph*{\bf Self-attention Network }
We integrate temporal features with spatial means through element-wise addition denoted as $\textbf{x} \oplus \pmb{\rho}$. Our approach involves the utilization of a three-layer multi-head attention network parameterized as $\Phi$, where the resulting features are considered as our spatial-temporal representations $\pmb{z}$. These representations are employed to enhance the close-form solution in Eq.~\eqref{eq: spaercloseform}, facilitating the generation of aligned images. This module ensures the incorporation of both temporal and spatial information for improved performance.
\paragraph*{\bf Geometric Deformation Correction}
We also develop a diffeomorphic deformation estimation model to guide our motion correction network, particularly addressing local distortions. Employing a Unet-based predictive deformable image registration neural network, we explore two theoretical developments: large deformation diffeomorphic metric mapping (LDDMM) that employs time-varying velocity fields~\cite{beg2005computing,hinkle2018lagomorph} and image registration using stationary velocity fields~\cite{arsigny2006log,balakrishnan2019voxelmorph}. The general formulation of the loss function is presented,
\begin{align}
\label{eq: lossgeo} 
l_{\text{Geo}} =  \sum^{T}_{t=0} \text{Dsim}[I_t \circ Q_t \circ \psi_t , I_{t+1}] + \text {reg} (\psi_t), 
\end{align}
where an interpolation operator $\circ$ deforms image $I_t \circ Q_t$ with an estimated transformation $\psi_t$. Here, $\text{Dsim}(\cdot, \cdot)$ measures the distance between the deformed and the target. The $\text{reg} (\cdot)$ is a regularization term that produces differentiable, bijective mappings with differentiable inverses. Such mappings in the space of diffeomorphisms highlight a set of desirable features, which have demonstrated noteworthy improvements in various medical image-related tasks~\cite{wang2023metamorph,wang2022geo}.
\paragraph*{\bf Network loss}
Combing all the modules we developed above, we write out the total loss for network training,
\begin{align}
\label{eq: loss} 
\ell &= \sum^{T}_{t=0}  \bigl\{ \text{Dist} \bigl\{ I_t \circ Q_t[\Phi(\Theta , \rho_t), \Phi(\Theta , \rho_{t+1})], \  I_{t+1} \bigl\}  \bigl\}   + \beta \cdot l_{\text{Geo}} + \textbf{Reg} (\Theta, \Phi) ,
\end{align}
where $\beta$ is a weighting factor balancing the contributions of both losses. The $\textbf{Reg}(\cdot)$ is a regularization term constrained on network parameters.
\begin{figure}[!bt]
\centering
 \includegraphics[width=.7600\textwidth] {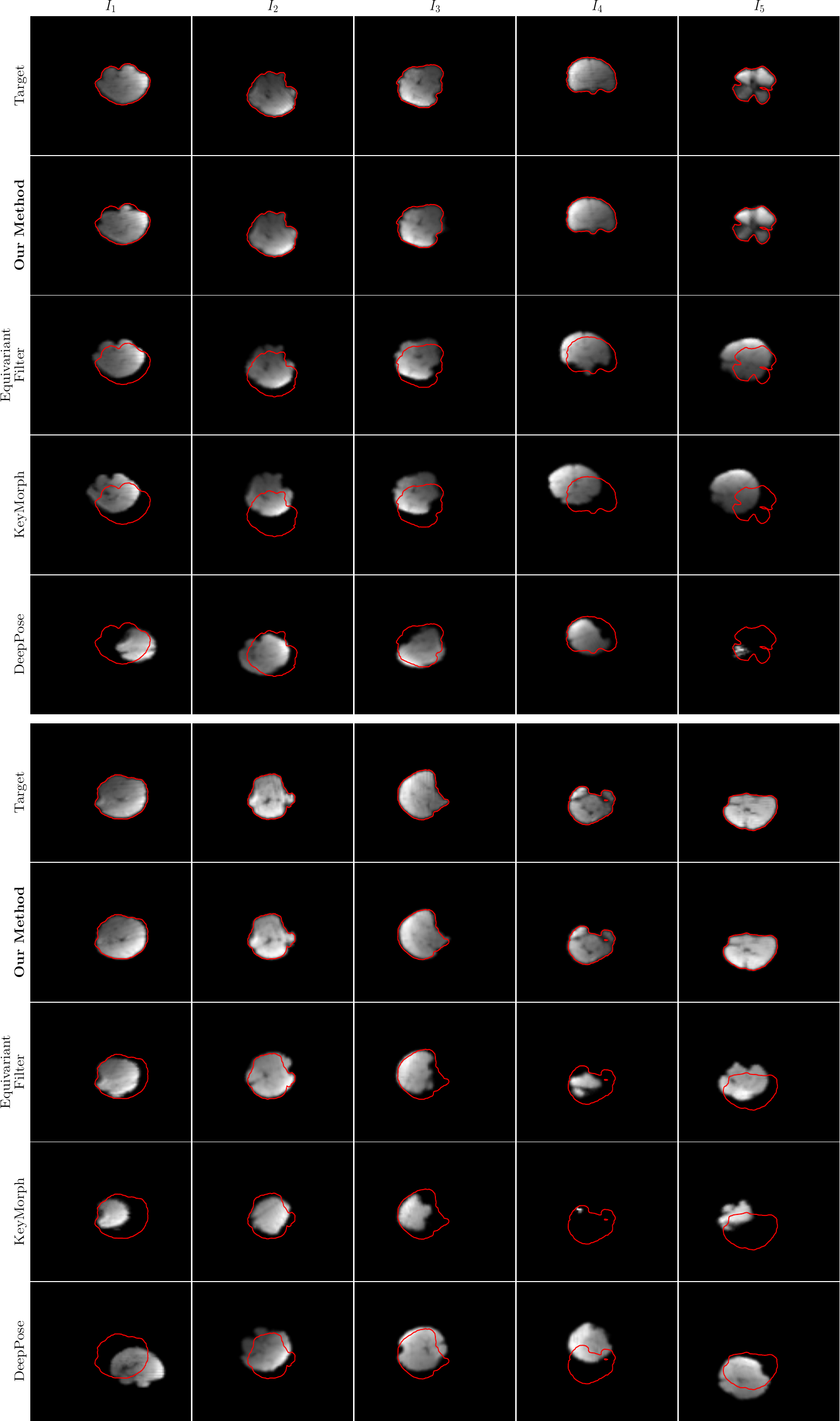}
 \caption{Two case studies (top and bottom half) serve as visualizations of motion tracking results, with the "target" fetal brains highlighted by red contours for all methods. Top to bottom for each case, target, motion-corrected results using our method, Equivariant filter~\cite{moyer2021equivariant}, KeyMorph\cite{evan2022keymorph} and DeepPose~\cite{salehi2018real}. }
\label{fig:example}            
\end{figure}
\section{Experimental Evaluation}

\paragraph*{\bf Data.}
We used $240$ sequences of 4D EPIs from fMRI time series of $15$ subjects who underwent fetal MRI scans (Siemens 3T scanner). The study was approved by the institutional review board and written informed consent was obtained from all participants. The dataset covers gestational ages from $22.57$ to $38.14$ weeks (mean $32.39$ weeks). Imaging parameters included a slice thickness of $2$ to $3 mm$, a repetition time (TR) of $2$ to $5.6$ seconds (mean $3.1$ seconds), an echo time (TE) of $0.03$ to $0.08$ seconds (mean $0.04$ seconds), and a flip angle (FA) of $90$ degrees. Fetal brains were extracted from scans using a real-time deep learning segmentation method~\cite{faghihpirayesh2022deep}. All brain scans were resampled to $96^3$ with a voxel resolution of $3 mm^3$ and underwent intensity normalization.

\paragraph*{\bf Baselines \& Evaluation Metrics.}
We compared the SpaER model with static learning-based motion correction methods, DeepPose~\cite{singh2020deep}, KeyMorph~\cite{evan2022keymorph}, and Equivariant Filters~\cite{moyer2021equivariant}. This included visual and quantitative analyses of translational and angular errors in simulated motion data. We also tested all models on spatial-temporal sequences from fMRIs with real fetal movements by manually adjusting translations and rotations. The effectiveness and stability of our model were highlighted through a Dice coefficient analysis, assessing alignment accuracy across varying motion degrees and sequence lengths. Our code is released online, \url{https://github.com/IntelligentImaging/SpaER}

\paragraph*{\bf Implementation \& Parameters} 
For model training, we used an initial learning rate of $\eta = 1e-5$ with the Adam optimizer and a batch size of 4. We employed cosine annealing for the learning rate schedule. The model parameters were set to $d = 96$ and $\beta = 0.5$. The dataset was split into $70\%$ training (176 sequences from 11 subjects), $15\%$ validation (32 sequences from 2 subjects), and $15\%$ testing (32 sequences from 2 subjects). The best-performing networks were saved based on validation performance across all models. Experiments were conducted on an NVIDIA RTX A6000 GPU.

\section{Results}
Table~\ref{tab} presents the motion correction errors (both translational and angular) comparison across all models for simulated motions. Our method produces the lowest errors ($\sim$ 3.8 mm of movement for the fetal brain) with the lowest variance between adjacent 3D volumes, indicating the best motion tracking accuracy.  
\begin{table}[!bt]
 \caption{Motion correction comparison of translation and rotation errors of various methods on 70 sequences of real fetal fMRI scans with simulated motions.} 
    \centering
    \begin{tabular}{c|>{\columncolor{red!30}}c|ccc}
      \hline
      Metric/Methods & Our method & Equivariant filter & KeyMorph & DeepPose \\ \hline
      Translation Error (mm) & \textbf{3.81 ± 1.05} & 9.42 ± 2.94 & 15.75 ± 3.90 & 21.99 ± 3.81 \\ \hline
      Angular Error (degree) & \textbf{2.76 ± 1.13} & 6.32 ± 2.89 & 9.08 ± 3.31 & 9.14 ± 3.29 \\ \hline
      Data Augmentation & No & No & Yes & Yes \\ \hline
      Spatial-temporal & Yes & No & No & No \\ \hline
    \end{tabular}
    \label{tab}
\end{table}
\begin{figure}[!bt]
\centering
\includegraphics[width=1.0\linewidth]{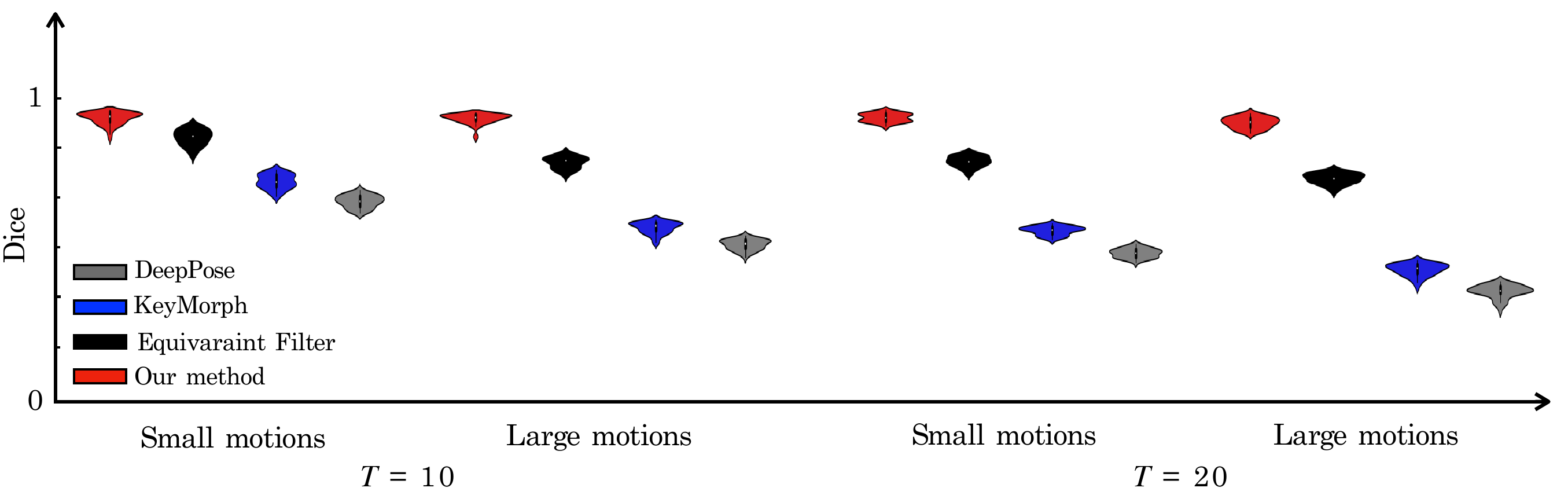}
     \caption{Motion tracking performance in real fMRI across varying degrees and lengths of motion sequences ($T$). Small ($\mathcal{T}{max} = 10 \text{mm}$, $\mathcal{R}{max} = 5^{\circ}$) and large motions ($\mathcal{T}{max} = 30 \text{mm}$, $\mathcal{R}{max} = 20^{\circ}$) were evaluated. Report efficiency with average time consumption: \textbf{0.501s} per pair / \textbf{9.960s} per sequence when $T=20$.}
\label{fig:dice}            
\end{figure}

Figure~\ref{fig:dice} quantitatively shows a model accuracy comparison over varying degrees of motions and different lengths of data sequences. Our model exhibits superiority in handling real motions ranging from small to large, and it maintains comparable motion tracking accuracy when dealing with extended data sequences. This indicates the high stability and robustness of our model, as it demonstrates a high level of accuracy in correcting severe fetal brain movements. We also report the average time consumption for adjacent pairs and the entire sequence. The total computation time of our model for motion tracking is 10 seconds for a 4D sequence that takes $\sim60s$ to acquire. This paves the way for an efficient real-time tracking of the fetal head motion for prospective correction.

Figure~\ref{fig:example} visualizes the results of motion tracking for two representative cases for all methods. Our model outperforms other baselines by producing the best alignment results with negligible errors. This demonstrates the effectiveness of our model in correcting rigid motions and geometric distortions.

\section{Conclusion}
This paper presents a novel model for fetal motion tracking via learning spatio-temporal equivariant representations. In contrast to existing approaches that estimate motion parameters at individual time frames, our approach takes a temporal perspective on the motion pattern of the fetal brain, achieving this by integrating time information into the low-dimensional spatial means of the image data. This framework excels in achieving the best and most stable performance across unpredictable fetal motions and excessive movement of fetuses, both in simulated and real scenarios, showcasing great potential in real-time tracking and steering for fetal head monitoring systems. A promising avenue for future research stemming from this study involves applying our trained model to various image modalities for motion correction tasks, all without requiring network retraining, as it inherently learns the characteristics of rigid transformations.
\section{Acknowledgments} This research was supported in part by the National Institute of Biomedical Imaging and Bioengineering, the National Institute of Neurological Disorders and Stroke, and Eunice Kennedy Shriver National Institute of Child Health and Human Development of the National Institutes of Health (NIH) under award numbers R01NS106030, R01EB031849, R01EB032366, and R01HD109395; and in part by the Office of the Director of the NIH under award number S10OD025111. This research was also partly supported by NVIDIA Corporation and utilized NVIDIA RTX A6000 and RTX A5000 GPUs. The content of this publication is solely the responsibility of the authors and does not necessarily represent the official views of the NIH or NVIDIA.
\bibliographystyle{splncs04}
\bibliography{miccai2024references}

\end{document}